\RequirePackage[2020-02-02]{latexrelease}
\documentclass[aps,twocolumn,
10pt,aps,amsmath,amssymb,nofootinbib,superscriptaddress]{revtex4}
\usepackage{braket}
\usepackage{sans}
\usepackage{hyperref}
\usepackage{newtxtext,newtxmath}
\usepackage{tikz}

\usepackage{booktabs,multirow,tabularx}
\usepackage{blindtext}
\usepackage{float}
\usepackage{physics}
\usepackage{url}
\usepackage{eqnarray,amsmath}
\usepackage[export]{adjustbox}
\usepackage{graphicx}
\usepackage{bbold}
\usepackage{ amssymb }
\begin{document}
 \title{Neutrino spin-flavour precession in magnetized white dwarf}
 	\author{Jyotismita Adhikary}
	\email{adhikary.2@iitj.ac.in}
	\affiliation{Indian Institute of Technology Jodhpur, Jodhpur 342037, India}
	\author{Ashutosh Kumar Alok}
	\email{akalok@iitj.ac.in}
	\affiliation{Indian Institute of Technology Jodhpur, Jodhpur 342037, India}
	\author{Arindam Mandal}
	\email{mandal.3@iitj.ac.in}
	\affiliation{Indian Institute of Technology Jodhpur, Jodhpur 342037, India}
	\author{Trisha Sarkar}
	\email{sarkar.2@iitj.ac.in}
	\affiliation{Indian Institute of Technology Jodhpur, Jodhpur 342037, India}
	\author{Shreya Sharma}
	\email{sharma.81@iitj.ac.in}
	\affiliation{Indian Institute of Technology Jodhpur, Jodhpur 342037, India}
 \begin{abstract}
    Due to  notoriously small value of the neutrino magnetic moment, the phenomena of neutrino spin flavour precession  (SFP) requires very high magnetic field. This makes only a handful of systems suitable to study this phenomena. By the observation of SFP, the Dirac and Majorana nature of neutrinos is expected to be distinguished. In this work, we point out the potential of white dwarf (WD) system in studying the spin-flavour oscillation of neutrinos. From recent analysis, it has been found that young isolated WDs may harbor very strong internal magnetic field, even without exhibiting any surface magnetic field. The presence of magnetic field enhances the cooling process and along with that, renders the spin-flavour oscillation of neutrinos emitted in the neutrino cooling process. Employing the standard WD specifications, we analyse whether a magnetized WD is a suitable environment to distinguish between the Dirac and Majorana nature of neutrino. Lower value of spin flavour transition probability implies reduced active neutrino flux which is possible to be estimated in terrestrial neutrino detectors. We find that the spin flavour transition probability of Dirac neutrinos is much higher in comparison to the Majorana neutrino which converts the active neutrino flavours to sterile in a significant amount. We also examine the sensitivity of the spin flavour transition probability to the neutrino magnetic moment.
    
\end{abstract}
\maketitle
\section{Introduction}\label{intro}
The current understanding of fundamental interaction of nature is inculcated in a theory known as standard model (SM) of electroweak interactions. Despite its baronial success, it cannot be considered as the ultimate theory  of fundamental interactions of nature. This is because SM fails to explain the observed matter-antimatter asymmetry of the universe, the origin of dark matter and dark energy. Therefore one needs to hunt for beyond SM physics. Neutrinos are the mysterious fundamental particles which has the potential to reveal many of the deep kept secrets of the universe. 
  Electromagnetic properties of neutrinos is an excellent probe to explore the domain of such new physics (NP) \cite{Giunti:2008ve, Studenikin:2019ggv, Broggini:2012df, Giunti:2014ixa}.
In SM, neutrinos are massless Weyl fermions which are electrically neutral. They undergo weak interaction with SM quarks and leptons and lead to the conservation of individual lepton numbers which is an accidental symmetry in the SM. Weyl neutrinos do not usually couple with the magnetic field due to their vanishing mass. However, the phenomena of neutrino oscillation manifests the fact that neutrinos are massive. Presence of neutrino mass may result in various extensions of SM described by different models \cite{PhysRevLett.58.1807}

In minimally extended SM (MESM) neutrinos may gain both mass and electromagnetic (EM) properties. 
The simplest MESM operates under the gauge group $SU(3)_c\times SU(2)_{L}\times SU(2)_R \times U(1)_{B-L}$. Models based on these gauge groups are left-right symmetric (LRSM) and involve parity invariance which is violated in standard electroweak (EW) theory. 
 One of the fundamental question in particle physics is whether the neutrinos are Dirac or Majorana in nature \cite{Bilenky:2020vjk} 
 on which EM properties of neutrinos can shed some light due to the difference in form factors associated with Dirac and Majorana neutrinos. In MESM, after EW symmetry breaking, if a neutrino gains Majorana mass, an antineutrino with equal mass is to be added which can couple at a higher order with the magnetic field through radiative corrections via quantum loop effect, while if the neutrino acquires Dirac mass a right handed heavy neutrino is to be added to the SM fermion spectra for undergoing seesaw mechanism \cite{Giunti:2008ve, Masood:2015pha}.
In MESM, neutrino carries very small magnetic moment ($\leq 10^{-19}~\mu_{B}$), while the value predicted from several NP models are quite larger than the MESM estimation ($10^{-12}-10^{-10}~\mu_B$). Different astrophysical and cosmological observations and the results obtained from various terrestrial experiments like GEMMA, XENON etc can constrain the value of neutrino magnetic moment. The requirement of large neutrino mass for its large magnetic moment poses a problem as current limit of neutrino mass is extremely small. However, this issue is resolved in several theoretically predicted models .

One of the most significant consequence of neutrino EM properties is the phenomenon of neutrino spin oscillation in presence of external magnetic field which can affect various astrophysical phenomena observable in terrestrial experiments. Dense astrophysical compact objects with huge magnetic field are significant for the analysis of spin flavour precession (SFP). In our work we have chosen WD which are the most abundant stars currently present in the globular clusters in our Milky Way galaxy. WDs are unique compact stellar objects which are considered to be "dead stars" being the products of stellar evolution of its progenitor star formed at its end stage \cite{Balberg:2000xu}. The stellar structure is sustained by hot plasma containing degenerate electrons and neutrinos are produced in copious amount in WDs during its cooling phase via different mechanisms which can then possibly undergo the phenomena of SFP. The presence of magnetic field affects various properties of WD stars including its thermal evolution alongwith its neutrino emission \cite{Peterson:2021teb}. In this work, we study the SFP phenomena of neutrino in WD interior induced by their large magnetic field, considering the scenario of both Dirac and Majorana neutrinos. We find that the spin flavour transition probability at the surface of WD is extremely sensitive to 
the value of the magnetic moment taken into consideration. Also the Dirac and Majorana nature of neutrinos can be distinguished successfully. 

The plan of the work is as follows. In section \ref{form},  we provide the formalism of our work explaining briefly about the neutrino magnetic moment and its consequence in the compact star. 
Then we present our results in section \ref{res} and conclude finally in section \ref{con}.
\section{Formalism}\label{form}
In this section we describe the theoretical framework of our analysis. In subsection \ref{form1}, we provide a brief review of the neutrino magnetic moment, in subsection \ref{form2} we demonstrate the formalism of neutrino spin flavour precession phenomena, while in subsection \ref{form3} we discuss about its application in WD star.

\subsection{Neutrino magnetic moment}\label{form1}
It is well known that within the framework of SM, neutrinos are massless which results in zero magnetic moment.  Below the SM symmetry breaking scale the neutrino gains mass and consequently an effective magnetic moment. The operators containing  neutrino magnetic moment and mass terms are given by \cite{Babu:2020ivd}
\begin{equation}\label{eq}
\mathcal{L} \subset \mu_{\nu}\bar{\nu}_L \sigma_{\mu\nu}\nu_R F^{\mu\nu}+m_{\nu}\bar{\nu}_L\nu_R+h.c.
\end{equation} 
Here $\mu_{\nu}$ is the neutrino magnetic moment and $\sigma_{\mu\nu}$ is the corresponding Lorentz tensor operator which is antisymmetric, $\sigma_{\mu\nu}=[\gamma_{\mu},\gamma_{\nu}]$ while the mass term is symmetric being a Lorentz scalar. $F^{\mu\nu}$ denotes the EM field tensor. The dimension of the two kinds of operators depend on whether the NP scale is above or below the EW scale \cite{Lindner:2017uvt}.

 In MESM including right handed neutrinos, the magnetic moment of Dirac neutrinos are diagonal and is given by \cite{Grigoriev:2018cvo, Yilmaz:2016ilw, Fujikawa:1980yx}
\begin{equation}\label{eq1}
\mu_{ii}^{D}=\frac{3eG_F m_i}{8\sqrt{2}\pi^2}\approx 3.2\times 10^{-19}\frac{m_i}{1eV} \mu_B.
\end{equation}
For Majorana neutrinos the diagonal magnetic moments vanish and transition magnetic moments are only non zero ($\mu_{ij}^{M},i\neq j$), given by \cite{PhysRevD.25.766}
\begin{equation}\label{eq1a}
\mu_{ij}^{M}=-\frac{3eG_F}{32\sqrt{2}\pi^2}(m_i \pm m_j)
\sum_{l=e,\mu,\tau} U_{li}^{*} U_{lj} \frac{m_{l}^{2}}{m_{W}^{2}}.
\end{equation}
In Eq. \eqref{eq1} and \eqref{eq1a} $\mu_B$ and $m_i$ denote Bohr magneton and neutrino mass, while $e$ and $G_F$ represent the electronic charge and Fermi constant, respectively, $m_l$ is the mass of $l$-th charged lepton and $m_W$ is the mass of $W$-boson. This can explicitly distinguish between the Dirac and Majorana nature of neutrino which is caused by the different form factors in case of the two kinds of neutrinos. In case of massless Weyl neutrinos the presence of magnetic moment is forbidden due to the absence of right handed neutrino state. Neutrino magnetic moment does not appear at tree level, while higher order perturbative expansion can incorporate it which can be understood by some effective vertex. 
The transition dipole moments are much smaller than the diagonal magnetic moments due to GIM mechanism \cite{Giunti:2008ve, Xing:2012gd}. 
It was first suggested by Voloshin {\it{et al.}} \cite{Voloshin:1986ty} that large magnetic moment of neutrino is required to explain the SNP anomaly which can be acheived by the presence of a charged singlet scalar particle \cite{Fukugita:1987ti, Rajpoot:1991fa}, capable of coupling to quarks \cite{Babu:1993hx}.

From Eqn. (1) it can be shown that large magnetic moment ($\sim 10^{-10}~\mu_B$) is expected if neutrinos possess larger mass. However, there are several models which involve additional symmetries alongwith SM which provide sufficiently larger magnetic moments even with suppressed  neutrino masses \cite{Babu:2020ivd,Babu:1992vq, Barr:1990um, PhysRevLett.63.228}.

 Current limit on the neutrino magnetic moment is about $\sim 10^{-10}-10^{-12}~\mu_{B}$ \cite{Grigoriev:2018cvo, Beda:2012zz,Borexino:2017fbd} which is much greater than the theoretically predicted value obtained from MESM ($\sim 10^{-19}~\mu_{B}$) \cite{Broggini:2012df}. 
Different NP models provide large enhancement in the values of the neutrino magnetic moments \cite{Ayala:1998qz, Joshipura:2002bp, Beda:2009kx, Canas:2015yoa, Grimus:2002vb, Babu:2020ivd}. These models include minimal supersymmetric models (MSSM) \cite{Aboubrahim:2013yfa, Fukuyama:2003uz}. In a model independent framework, non standard interaction (NSI) of neutrinos can also constrain the range of the magnetic moment \cite{Giunti:2014ixa, Healey:2013vka, Papoulias:2015iga, PhysRevD.103.095004}.

Astronomical measurements such as the stellar evolution can place constraints on the magnetic moment indirectly. Among several techniques,
 the observation of the tip of the red giant branch in globular clusters is a significant method of restricting the range of magnetic moment providing an average value of $10^{-12}~\mu_B$ \cite{Capozzi:2020cbu, PhysRevLett.111.231301, 2015APh....70....1A} which is an order smaller as compared to the limit obtained by the terrestrial experiments. Reactor experiments like GEMMA, XENON and accelerator experimental setups such as Borexino, LSND involve the process of electron neutrino elastic scattering which is possible to be mediated by photons due to the nonzero neutrino magnetic moment, to put constraints on the magnetic moment by measuring its differential scattering cross section. From a recent analysis based on XENON1T experiment, the transition magnetic moment $\mu_{e\mu}$ is found to be existing in the range $1.65-3.42 \times 10^{-11}~ \mu_B$ \cite{Babu:2020ivd} which is larger than any experimental or theoretical predictions till now. Electromagnetic properties of neutrino exhibit several significant processes such as spin light, neutrino decay, Cherenkov radiation, spin flip phenomena etc. In this work we tend to focus on the spin flip phenomena which is discussed in the next section in brief.

\subsection{Phenomena of neutrino spin-flavour precession}\label{form2}
In the presence of an external magnetic field if the neutrino possesses a nonzero magnetic moment, 
the magnetic field present in the medium causes a precession of the neutrino spin. This results in the mixing between the left handed (LH) and right handed (RH) neutrinos which provides another possible explanation for the deficiency observed in solar neutrino flux \cite{Okun:1986hi}, beside the occurrence of neutrino flavour oscillation. The phenomenon of neutrino spin oscillation in addition with its flavour mixing has been studied previously \cite{Nunokawa:1993dr, Grimus:2002vb}.  

The time evolution of neutrino state is given by the Schrodinger equation. In two flavour scenario for Dirac neutrinos it is given as \cite{joshi2020neutrino}
\begin{widetext}
\begin{eqnarray}\label{eq2}
i\frac{d}{dt} \begin{pmatrix}
\nu_{eL} \\
\nu_{\mu L}\\
\nu_{eR}\\
\nu_{\mu R}
\end{pmatrix} = H_D\begin{pmatrix}
\nu_{eL} \\
\nu_{\mu L}\\
\nu_{eR}\\
\nu_{\mu R} 
\end{pmatrix}, 
H_D=\begin{pmatrix}
\frac{-\Delta m^{2}}{4E_{\nu}}cos 2\theta+V_e & \frac{\Delta m^{2}}{4E_{\nu}}sin 2\theta & \mu_{ee}B_\perp & \mu_{e\mu}B_\perp\\
\frac{\Delta m^{2}}{4E_{\nu}}sin 2\theta & \frac{\Delta m^{2}}{4E_{\nu}}cos 2\theta+V_\mu & V_{e\mu}B_\perp & \mu_{\mu\mu}B_\perp \\
\mu^{*}_{ee}B_\perp & \mu^{*}_{\mu e}B_\perp & \frac{-\Delta m^{2}}{4E_{\nu}}cos 2\theta & \frac{\Delta m^{2}}{4E_{\nu}}sin 2\theta \\
\mu^{*}_{e\mu}B_\perp & \mu^{*}_{\mu \mu}B_\perp & \frac{\Delta m^{2}}{4E_{\nu}}sin 2\theta  &  \frac{\Delta m^{2}}{4E_{\nu}}cos 2\theta \\ 
\end{pmatrix}
\end{eqnarray}
\end{widetext}
From \eqref{eq2} it can be seen that both diagonal and off-diagonal magnetic moments are allowed in case of Dirac neutrinos. 
In case of Majorana neutrinos eqn. \eqref{eq2} is modified as follows \cite{joshi2020neutrino}
\begin{widetext}
\begin{eqnarray}\label{eq3}
i\frac{d}{dt} \begin{pmatrix}
\nu_{eL} \\
\nu_{\mu L}\\
\bar{\nu}_{e R}\\
\bar{\nu}_{\mu R}
\end{pmatrix} = H_M\begin{pmatrix}
\nu_{eL} \\
\nu_{\mu L}\\
\bar{\nu}_{eR}\\
\bar{\nu}_{\mu R} 
\end{pmatrix},
H_M=\begin{pmatrix}
\frac{-\Delta m^{2}}{4E_{\nu}}cos 2\theta+V_e & \frac{\Delta m^{2}}{4E_{\nu}}sin 2\theta & 0 & \mu_{e\mu}B_\perp\\
\frac{\Delta m^{2}}{4E_{\nu}}sin 2\theta & \frac{\Delta m^{2}}{4E_{\nu}}cos 2\theta+V_\mu & -\mu_{e\mu}B_\perp & 0 \\
0 & -\mu^{*}_{\mu e}B_\perp & \frac{-\Delta m^{2}}{4E_{\nu}}cos 2\theta-V_e & \frac{\Delta m^{2}}{4E_{\nu}}sin 2\theta   \\
\mu^{*}_{e\mu}B_\perp & 0 &  \frac{\Delta m^{2}}{4E_{\nu}}sin 2\theta &  \frac{\Delta m^{2}}{4E_{\nu}}cos 2\theta-V_\mu  \\ 
\end{pmatrix}
\end{eqnarray}
\end{widetext}
In Majorana framework, only transition magnetic moments are possible. Here $\nu_{eR}-\nu_{\mu R}$ mixing can occur. In Eqn. \eqref{eq2} and \eqref{eq3}, 
$B_\perp$ and $\mu$ represent transverse magnetic field and neutrino magnetic moment respectively. Here, $\Delta m^{2}=m_{2}^2-m_{1}^2$. $V_{e}$ and $V_{\mu}$ are the potentials experienced by $\nu_e$ and $\nu_{\mu}$ respectively. Further, $V_e=\sqrt{2}G_F (n_e-n_n/2)$ and $V_\mu=-\sqrt{2}G_F n_n/2$, where $n_e$ and $n_n$ are electron and neutron number density respectively. In both cases the conversion of one flavour into another with different helicity states are caused by non-diagonal magnetic moments.

The transition probability for $\nu_{eL}\rightarrow\nu_{\mu R}$ \hspace{1 mm} ($\nu_{e}\rightarrow\overline{\nu}_{\mu}$ for Majorana neutrinos) is \cite{joshi2020neutrino, Giunti:2014ixa}
\begin{equation}
    P(r) = {\sin^{2}{\theta_{eff}}} {\sin^{2}{\left(\frac{\pi r}{L_{eff}}\right)}} ,
\end{equation}\\
where $r$ is the radial distance inside the WD mesured from its center. Also 
\begin{equation}
    \sin^{2}{\theta_{eff}}= \frac{4{H_{(D,M)14}}^2}{4{H_{(D,M)14}}^2+(H_{(D,M)11}-H_{(D,M)44})^{2}}
\end{equation}\\and
\begin{equation}
    L_{eff}=\frac{2\pi}{\sqrt{4{H_{(D,M)14}}^2+(H_{(D,M)11}-H_{(D,M)44})^{2}}}
\end{equation}

To obtain the transition probability of the order of unity, the amplitude of transition probability, blue ($sin^2\theta_{eff}$) must not be lower than $0.5$ and the length traversed by the neutrino in the medium must be $x\geq L_{eff}/2$. Depending on these two conditions, a critical magnetic field can be defined above which the probability of spin flavour transition is significant and for the transition $\nu_{eL}\rightarrow \nu_{\mu R}$ is expressed as, 
\begin{equation}
B_{cr}=\frac{1}{2\mu}(H_{(D,M)_{44}}-H_{(D,M)_{11}})
\end{equation}
In case of the magnetic moment value predicted from SM, in the astrophysical object $B_{cr}$ becomes extremely large.This $B_{cr}$ then becomes  even larger than magnetic field of the astrophysical object, which leads to extremely small spin flavour transition probability of neutrino. Presence of a strong matter potential reduces the effect of spin precesion even in presence of the magnetic field \cite{PhysRevD.37.1368}.

In case of astrophysical environments containing highly dense plasma the effect of SFP may strongly show experimental evidences. In the interior of dense matter object like supernova due to the presence of SFP, in case of Dirac neutrinos, the sterile states produced may convert back to active flavours and give rise to high energy anomalous events in neutrino detectors alongwith the usual supernova events \cite{Raffelt:1999tx, Lychkovskiy:2009pm}. 

In the domain of cosmology, the effective number of active neutrinos is affected due to the existence of SFP. In case of Dirac neutrinos the number of active flavours is found to be greater than three due to the consideration of RH neutrino which contradicts with the observation, while in the case of Majorana neutrino, this issue does not appear as the RH counterpart actually is its own anti particle which also affects the cosmological parameters such as neutrino decoupling temperature and the abundance of different particle species \cite{Vassh:2015yza}. 
Our work includes the analysis of SFP in another class of specific compact stars which is to be discussed in the next section.

\subsection{Application in white dwarf}\label{form3}
The analysis of SFP has been done in several astrophysical environments \cite{Yilmaz:2016ilw,joshi2020neutrino,popov2019neutrino,sasaki2021neutrino}. We conduct our analysis in the system of WD which is another favourable compact medium due to its large magnetic field and source of ample production of neutrinos. Beside this, at present in Milky Way galaxy, the globular clusters mostly contain low mass stars ($\leq 1M_{\odot}$) which end up as WDs undergoing stellar evolution. The massive stars have already evolved faster than the stars with lower mass and ended up as neutron star (NS) or black hole (BH) after experiencing supernova explosion \cite{Raffelt:1999tx}. Therefore from observational point of view WDs get much more importance. In addition, the WDs present in the globuler clusters is able to provide information about the formation of the galaxy and the rate of star formation.
Furthermore, it is to be mentioned that since the matter density is somewhat lower in WD compared to the other dense matter objects like NS, so that the matter suppression is much lower than that of NS. 

White dwarfs (WD) are the end stage product of a main sequence star having low to medium mass ($<8M_{\odot}$). After the nuclear fuel at its core is completely exhausted, the star expands to a red giant ,which sheds its outer layer with time as planetary nebulae and its core part forms a WD in which the gravitational collapse of the system is supported by the electron degeneracy pressure. Nearly $\sim97\%$ of the stars end up as WDs in our galaxy \cite{2020NatAs...4..690P}. WD core is usually consisted of helium, while massive WDs may contain heavier nuclei like carbon or oxygen at their core. Mass of a WD is bound by Chandrashekhar limit ($1.475 M_{\odot}$) with interior density of $\sim 10^6$ cm$^3$ which usually has a carbon core surrounded by helium layer with a hydrogen envelope. For non-relativistic degenerate electron gas considering a polytropic equation of state for WD dense matter, the mass radius ($M-R$) relation is given by \cite{1983bhwd.book.....S}
\begin{equation}
R=10,500~km~(0.6M_{\odot}/M)^{1/3}(2Y_e)^{5/3},
\end{equation}
while the central density is represented as \cite{1983bhwd.book.....S}
\begin{equation}
\rho_{core}=1.46\times10^6~g~cm^{-3}(M/0.6M_{\odot})^2(2Y_e)^{-5}.
\end{equation}
 Here $Y_e$ is the electron density fraction and $0.6M_{\odot}$ is the canonical mass of WD. After its formation, WD starts to cool down. At the initial stage, the cooling mechanism occurs via the emission of neutrinos from its core which are generated by the process of  plasmon decay, upto the temperature of $\sim 10^9$ K and is the main efficient mechanism for cooling, while photons are emitted from its surface to continue the cooling phenomena at the latter stage when the temperature goes below $\sim 10^9$ K when the neutrino emission becomes Boltzmann suppressed \cite{1969ApJ...155..221S, 1969ApJ...156.1021K, 1966ApJ...146..437V, Bedaque:2012mr}. 
 
 A small fraction of single isolated WDs possess magnetic field. The origin of the WD magnetic field is predicted to be generated from its progenitor star or from the evolution of a binary system. Magnetic field present in WDs may have large impact on the emitted neutrino flux. It is to be noted that the WDs having no magnetic field present on their surface may still have a large internal magnetic field hidden beneath the surface as their predecessor RG star also possesses large magnetic field \cite{2016ApJ...824...14C}.
 The nature of the field geometry is still unknown, hence in our analysis we have considered constant value of magnetic field throughout the star. From several asteroseismological analysis \cite{2015Sci...350..423F} and observations of spectral lines the magnetic field of the surface of the WDs is constrained, while the exact estimation of the magnetic field at the core is still uncertain. In a recent analysis, the predicted upper limit of the maximum surface magnetic field of a WD based on neutrino cooling of WD is given by \cite{Drewes:2021fjx}
 \begin{equation}
    B_{max}=8.8\times10^{11}G\left(\frac{M}{0.6M_{\odot}}\right)^{1/3}.
\end{equation} 
Presence of magnetic field affects the plasmon decay and hence the emission of neutrinos produced in the process. In presence of large magnetic field, the energy of the electrons are quantized and the electrons are confined in the ground state of different Landau levels, while at lower magnetic fields all the Landau levels are filled by the electrons and the quantization effect becomes redundant. In presence of large magnetic field the neutrino flux emitted from the WD is increased due to the production of the additional pair of neutrino emitted by the process of neutrino pair syncroton radiation which occurs due to the rotation of the electron around the magnetic field lines ($e\rightarrow e \nu \bar{\nu}$).

The neutrinos emitted from WD usually have their energy in MeV range. The detection of such neutrino flux at lower energy end is possible in many potential terrestrial detectors with large volumes \cite{Fraija:2019qpp, Abe:2011ts, LENA:2011ytb, JUNO:2015zny} like several water and ice Cherenkov detectors.


\section{Result and discussion}\label{res}
\begin{figure}
\begin{center}
\includegraphics[scale=0.3]{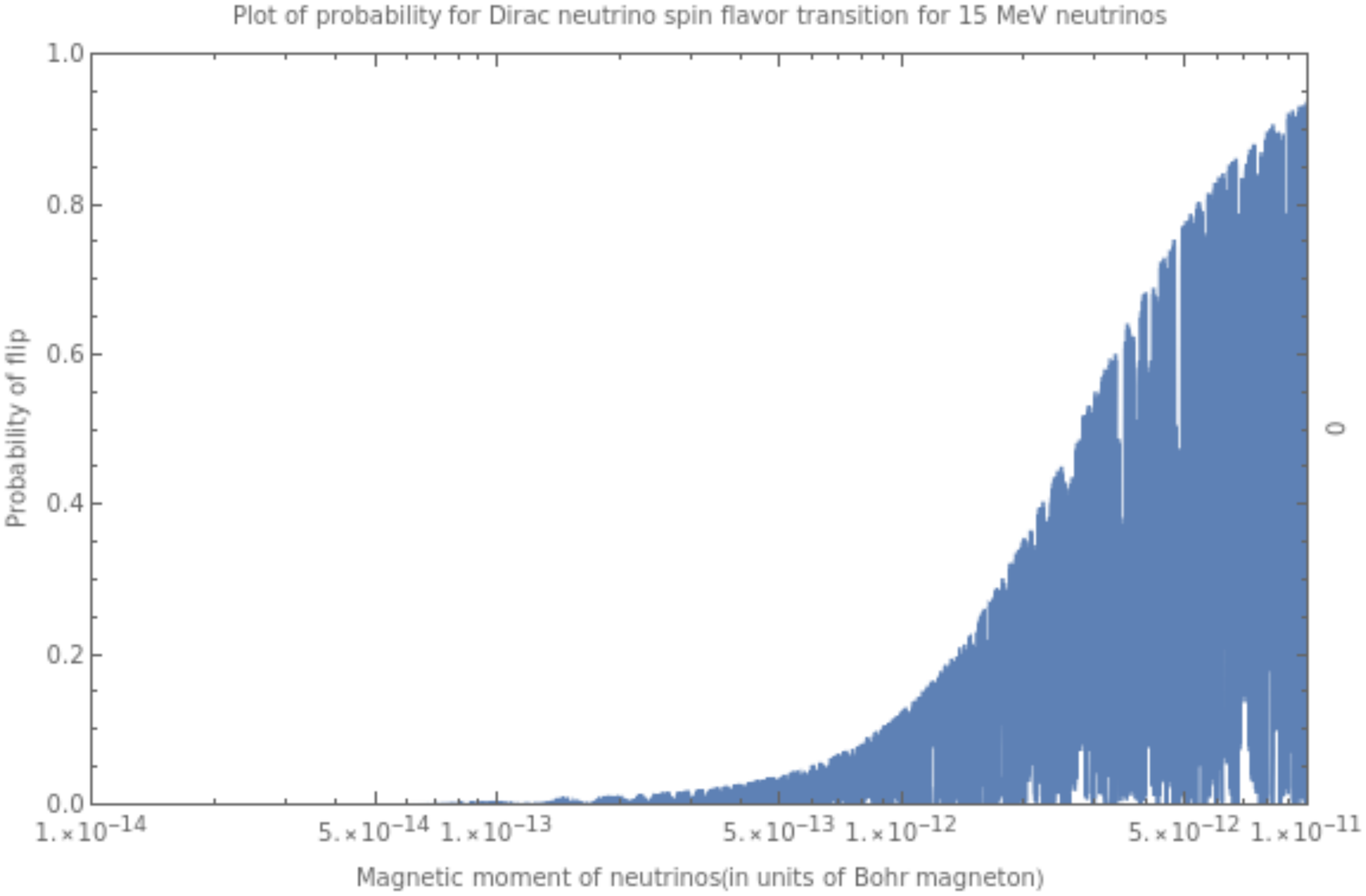}
\caption{Variation of Dirac neutrino spin flavour transition probability with magnetic moment for a particular neutrino energy of 15MeV}
\label{f1}
\end{center}
\end{figure}

\begin{figure}
\centering
\includegraphics[scale=0.3]{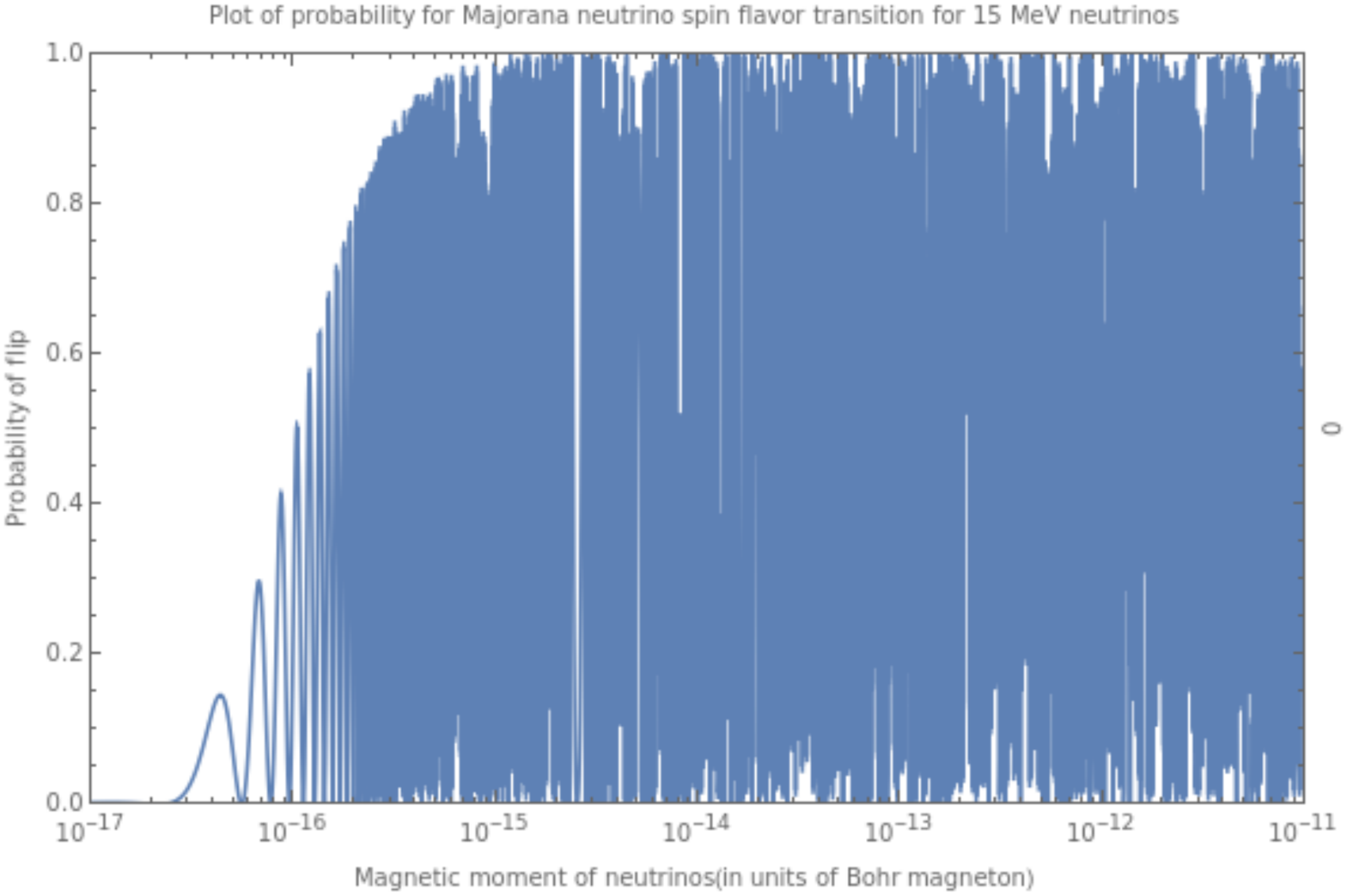}
\caption{Variation of Majorana neutrino spin flavour transition probability with magnetic moment for a particular neutrino energy of 15MeV}
\label{f2}
\end{figure}
In this section we describe the analysis of the neutrino SFP phenomenon inside the magnetized WD star considering both the scenarios of Dirac and Majorana neutrinos and discuss whether these two scenarios can be distinguished. We have chosen a comparatively young isolated WD composed of $npe$ matter with mass $0.6~M_{\odot}$ and radius $10,500$ km, possessing sufficiently high magnetic field of $8.8\times 10^{11}$ G for our analysis \cite{Drewes:2021fjx}. 
The values of standard neutrino oscillation parameters considered in our work are given in Table \ref{t1} which are determined from the global analysis of the data obtained from oscillation experiments 
 \cite{Esteban:2018azc}.
\begin{table}
\begin{center}
\begin{tabular}{ |c|c| } 
 \hline\hline
 Parameters & Best fit \\ 
 \hline
 $\theta_{12}^o$ & $33.82$ \\ 
 \hline
 $\Delta m_{12}^{2}\times10^{-5}$ eV$^2$ & $7.39$ \\ 
 \hline
\end{tabular}
\end{center}
\caption{Best fit values of standard neutrino oscillation parameters \cite{Esteban:2018azc}}
\label{t1}
\end{table}

In fig. \ref{f1} and fig. \ref{f2}
 we illustrate the plot showing the variation of spin flavour transition probability with neutrino magnetic moment for a fixed value of energy $15$ MeV in case of Dirac and Majorana neutrinos respectively. To generate all the plots electron and neutron fractions taken into are equal, $Y_e=Y_n=0.5$. In both the cases it can be observed that for extremely small value of magnetic moment ($\leq 10^{-19}~\mu_B$) which corresponds to the SM prediction, the transition probability tends to be zero. The probability remains non zero till the magnetic moment value is $\sim 6\times 10^{-14}\mu_B$ for Dirac while for Majorana it is $\sim 2.5\times 10^{-17}\mu_B$. From the two figures, it can be observed that the transition probability vary rapidly with very small change in neutrino magnetic moment.

\begin{figure}
\centering
\includegraphics[scale=0.3]{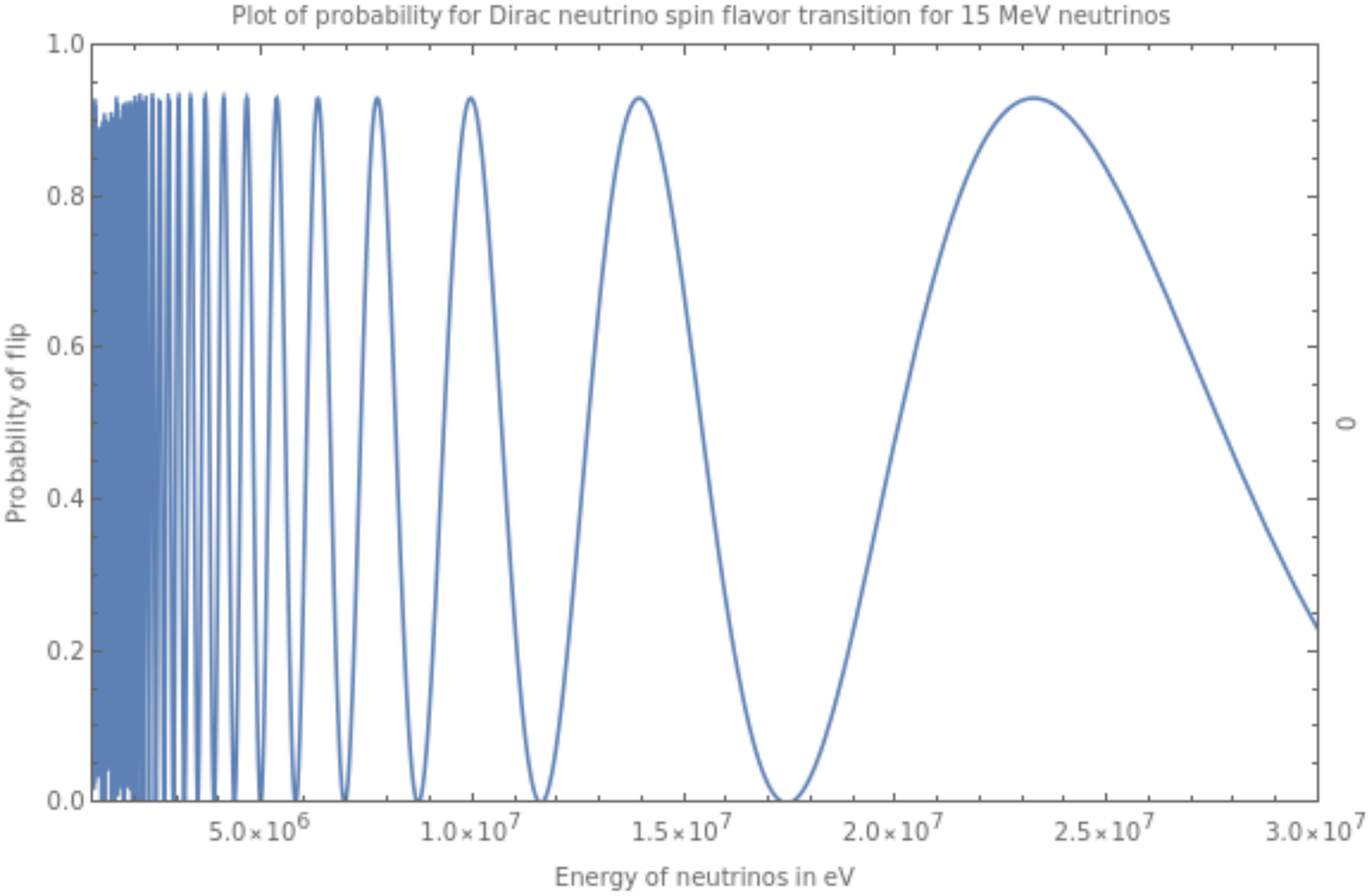}
\caption{Variation of Dirac neutrino spin flavour transition probability with energy}
\label{f3}
\end{figure}
 
\begin{figure}
\centering
\includegraphics[scale=0.3]{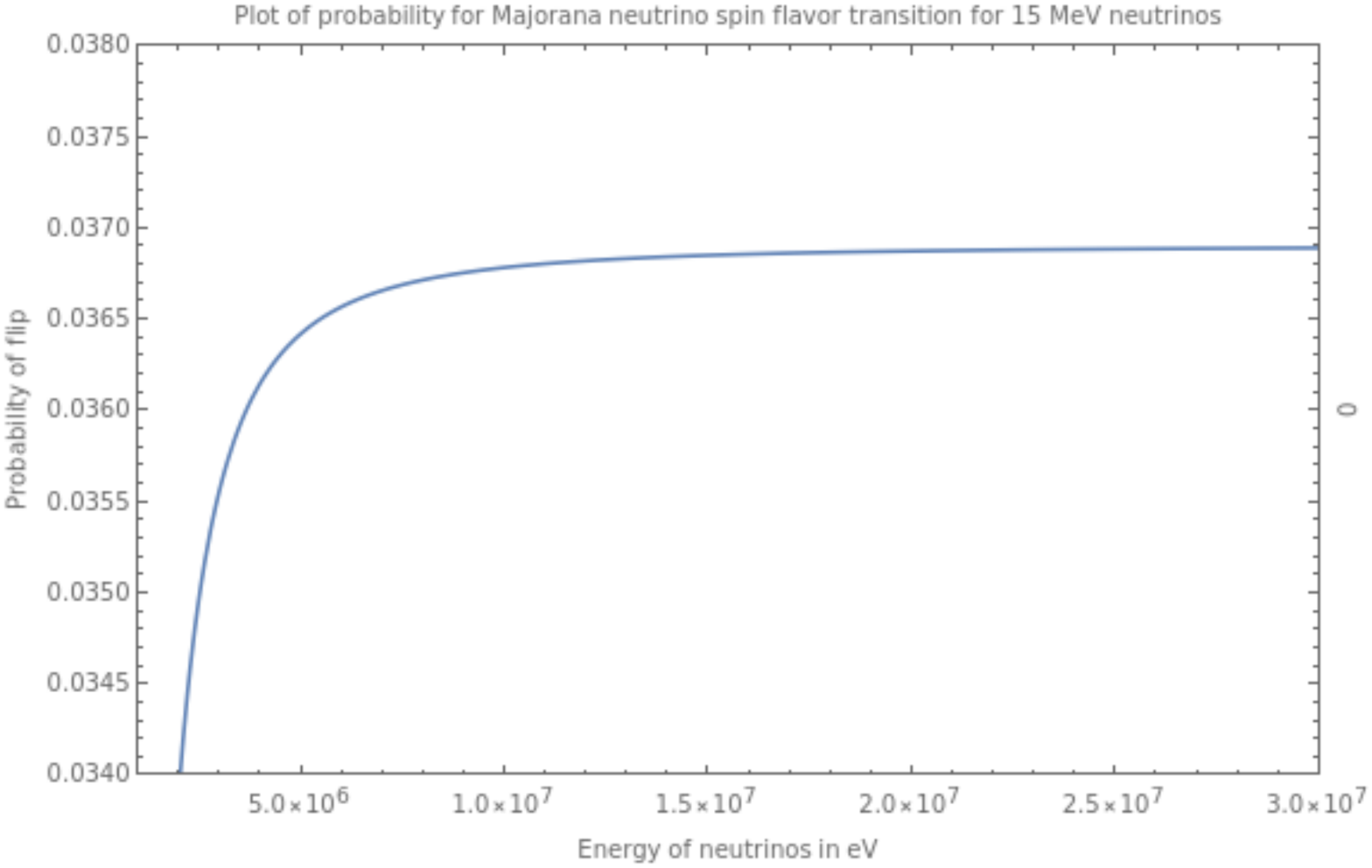}
\caption{Variation of Majorana neutrino spin flavour transition probability with energy}
\label{f4}
\end{figure} 
 
 

In fig. \ref{f3} and fig. \ref{f4} we illustrate the plot showing the variation of spin flavour transition probability with energy for a fixed value of neutrino magnetic moment $10^{-11}~\mu_B$ in case of Dirac and Majorana neutrinos respectively. The energy of the WD neutrinos vary usually upto approximately $30$ MeV due to which the enrgy scale is set accordingly. In case of Dirac neutrinos, the result shows rapid variation in transition probability while in case of Majorana, the result is drastically different which shows the probability is saturated to a nearly constant value of $\sim 0.037$.

From the pattern of the plot exhibited by fig. \ref{f3} and \ref{f4} one can easily distinguish Dirac and Majorana nature of neutrinos. All the results are obtained on the surface of the WD ($r\sim 10,500$ km).

In the case of Majorana neutrinos, the spin-flip probabilities saturates beyond few MeV energy of neutrinos. This is due to the cancellation of $V_e$ and $V_{\mu}$.

If the observed neutrino flux of a specific flavour at the detectors is perceived to be sufficiently reduced due to the active-sterile conversion then Dirac nature of neutrinos is concluded as one possible solution. On the other hand, if the flux of neutrino (anti neutrino) observed at the detector is much larger than its counterpart with opposite helicity, then it can be inferred that the neutrinos could possible be of Majorana nature.
  
\section{conclusion}\label{con}
We have considered the environment of a magnetic WD to study the phenomena of SFP of neutrino oscillation. The standard oscillation parameters are taken into account as obtained from the global analysis of data generated by oscillation experiments. The value of magnetic moment considered is chosen following different theretical and experimental predictions. We have considered a standard magnetized WD with nucleonic matter configuration having sufficiently large magnetic field with its value constrained from a recent analysis of neutrino cooling of a magnetic WD.

We find that magnetized WD is a convenient system which is able to distinguish between Dirac and Majorana nature of neutrino. We also observe that in Dirac picture, a significant fraction of neutrinos may may be converted to sterile and thus in detectors a reduction of incoming flux will be visible, while in case of Majorana framework, the probability of conversion of neutrinos to antineutrino is extremely small ($P \sim 0.037$) which implies no alteration of the incoming neutrino flux. Therefore, by proper and efficient detection technique the distinction between the two scenarios of neutrino can be established by the observation of the incoming neutrino flux.

Further, we have noticed that the transition probability is extremely sensitive to the choice of neutrino magnetic moment, in case of both Dirac and Majorana neutrino. The transition probability is found to be nearly vanishing for extremely small value of magnetic moment, the value of  which is found to be $\sim 10^{-13}~\mu_B$ in case of Dirac and $\sim 10^{-17}~\mu_B$ for Majorana neutrinos. 


\bibliographystyle{unsrt}
\bibliography{ref}

\begin{thebibliography}{10}

\bibitem{Giunti:2008ve}
Carlo Giunti and Alexander Studenikin.
\newblock {Neutrino electromagnetic properties}.
\newblock {\em Phys. Atom. Nucl.}, 72:2089--2125, 2009.

\bibitem{Studenikin:2019ggv}
Alexander Studenikin.
\newblock {Electromagnetic properties of neutrinos}.
\newblock {\em PoS}, EPS-HEP2019:374, 2020.

\bibitem{Broggini:2012df}
C.~Broggini, C.~Giunti, and A.~Studenikin.
\newblock {Electromagnetic Properties of Neutrinos}.
\newblock {\em Adv. High Energy Phys.}, 2012:459526, 2012.

\bibitem{Giunti:2014ixa}
Carlo Giunti and Alexander Studenikin.
\newblock {Neutrino electromagnetic interactions: a window to new physics}.
\newblock {\em Rev. Mod. Phys.}, 87:531, 2015.

\bibitem{PhysRevLett.58.1807}
M.~Fukugita and T.~Yanagida.
\newblock Particle-physics model for voloshin-vysotsky-okun solution to the
  solar-neutrino problem.
\newblock {\em Phys. Rev. Lett.}, 58:1807--1809, May 1987.

\bibitem{Bilenky:2020vjk}
S.~Bilenky.
\newblock {Neutrinos: Majorana or Dirac?}
\newblock 8 2020.

\bibitem{Masood:2015pha}
Samina~S. Masood.
\newblock {Magnetic Dipole Moment of Neutrino}.
\newblock {\em JHEP Grav. Cosmol.}, 1(1):56270, 2015.

\bibitem{Balberg:2000xu}
Shmuel Balberg and Stuart~L. Shapiro.
\newblock {The Properties of matter in white dwarfs and neutron stars}.
\newblock 4 2000.

\bibitem{Peterson:2021teb}
J.~Peterson, V.~Dexheimer, R.~Negreiros, and B.~G. Castanheira.
\newblock {Effects of Magnetic Fields in Hot White Dwarfs}.
\newblock {\em Astrophys. J.}, 921(1):1, 2021.

\bibitem{Babu:2020ivd}
K.~S. Babu, Sudip Jana, and Manfred Lindner.
\newblock {Large Neutrino Magnetic Moments in the Light of Recent Experiments}.
\newblock {\em JHEP}, 10:040, 2020.

\bibitem{Lindner:2017uvt}
Manfred Lindner, Branimir Radov\v{c}i\'c, and Johannes Welter.
\newblock {Revisiting Large Neutrino Magnetic Moments}.
\newblock {\em JHEP}, 07:139, 2017.

\bibitem{Grigoriev:2018cvo}
A.~Grigoriev, E.~Kupcheva, and A.~Ternov.
\newblock {Neutrino spin oscillations in polarized matter}.
\newblock {\em Phys. Lett. B}, 797:134861, 2019.

\bibitem{Yilmaz:2016ilw}
Deniz Yilmaz.
\newblock {Combined effect of NSI and SFP on solar electron neutrino
  oscillation}.
\newblock {\em Adv. High Energy Phys.}, 2016:1435191, 2016.

\bibitem{Fujikawa:1980yx}
Kazuo Fujikawa and Robert Shrock.
\newblock {The Magnetic Moment of a Massive Neutrino and Neutrino Spin
  Rotation}.
\newblock {\em Phys. Rev. Lett.}, 45:963, 1980.

\bibitem{PhysRevD.25.766}
Palash~B. Pal and Lincoln Wolfenstein.
\newblock Radiative decays of massive neutrinos.
\newblock {\em Phys. Rev. D}, 25:766--773, Feb 1982.

\bibitem{Xing:2012gd}
Zhi-zhong Xing and Ye-Ling Zhou.
\newblock {Enhanced Electromagnetic Transition Dipole Moments and Radiative
  Decays of Massive Neutrinos due to the Seesaw-induced Non-unitary Effects}.
\newblock {\em Phys. Lett. B}, 715:178--182, 2012.

\bibitem{Voloshin:1986ty}
M.~B. Voloshin and M.~I. Vysotsky.
\newblock {Neutrino Magnetic Moment and Time Variation of Solar Neutrino Flux}.
\newblock {\em Sov. J. Nucl. Phys.}, 44:544, 1986.

\bibitem{Fukugita:1987ti}
M.~Fukugita and T.~Yanagida.
\newblock {A Particle Physics Model for Voloshin-Vysotskii-Okun Solution to the
  Solar Neutrino Problem}.
\newblock {\em Phys. Rev. Lett.}, 58:1807, 1987.

\bibitem{Rajpoot:1991fa}
S.~Rajpoot.
\newblock {The Voloshin-Vysotskii-Okun solution to the solar neutrino problem
  in a left-right model}.
\newblock {\em Z. Phys. C}, 50:43--46, 1991.

\bibitem{Babu:1993hx}
K.~S. Babu, Kazuo Fujikawa, and Atsushi Yamada.
\newblock {Constraints on left-right symmetric models from the process b
  ---\ensuremath{>} s gamma}.
\newblock {\em Phys. Lett. B}, 333:196--201, 1994.

\bibitem{Babu:1992vq}
K.~S. Babu, D.~Chang, Wai-Yee Keung, and I.~Phillips.
\newblock {Comment on `Mechanism for large neutrino magnetic moments'}.
\newblock {\em Phys. Rev. D}, 46:2268--2269, 1992.

\bibitem{Barr:1990um}
Stephen~M. Barr, E.~M. Freire, and A.~Zee.
\newblock {A Mechanism for large neutrino magnetic moments}.
\newblock {\em Phys. Rev. Lett.}, 65:2626--2629, 1990.

\bibitem{PhysRevLett.63.228}
K.~S. Babu and R.~N. Mohapatra.
\newblock Model for large transition magnetic moment of the electron neutrino.
\newblock {\em Phys. Rev. Lett.}, 63:228--231, Jul 1989.

\bibitem{Beda:2012zz}
A.~G. Beda, V.~B. Brudanin, V.~G. Egorov, D.~V. Medvedev, V.~S. Pogosov, M.~V.
  Shirchenko, and A.~S. Starostin.
\newblock {The results of search for the neutrino magnetic moment in GEMMA
  experiment}.
\newblock {\em Adv. High Energy Phys.}, 2012:350150, 2012.

\bibitem{Borexino:2017fbd}
M.~Agostini et~al.
\newblock {Limiting neutrino magnetic moments with Borexino Phase-II solar
  neutrino data}.
\newblock {\em Phys. Rev. D}, 96(9):091103, 2017.

\bibitem{Ayala:1998qz}
Alejandro Ayala, Juan~Carlos D'Olivo, and Manuel Torres.
\newblock {Bound on the neutrino magnetic moment from chirality flip in
  supernovae}.
\newblock {\em Phys. Rev. D}, 59:111901, 1999.

\bibitem{Joshipura:2002bp}
Anjan~S. Joshipura and Subhendra Mohanty.
\newblock {Bounds on neutrino magnetic moment tensor from solar neutrinos}.
\newblock {\em Phys. Rev. D}, 66:012003, 2002.

\bibitem{Beda:2009kx}
A.~G. Beda, E.~V. Demidova, A.~S. Starostin, V.~B. Brudanin, V.~G. Egorov,
  D.~V. Medvedev, M.~V. Shirchenko, and Ts. Vylov.
\newblock {GEMMA experiment: Three years of the search for the neutrino
  magnetic moment}.
\newblock {\em Phys. Part. Nucl. Lett.}, 7:406--409, 2010.

\bibitem{Canas:2015yoa}
B.~C. Canas, O.~G. Miranda, A.~Parada, M.~Tortola, and Jose W.~F. Valle.
\newblock {Updating neutrino magnetic moment constraints}.
\newblock {\em Phys. Lett. B}, 753:191--198, 2016.
\newblock [Addendum: Phys.Lett.B 757, 568--568 (2016)].

\bibitem{Grimus:2002vb}
W.~Grimus, M.~Maltoni, T.~Schwetz, M.~A. Tortola, and J.~W.~F. Valle.
\newblock {Constraining Majorana neutrino electromagnetic properties from the
  LMA-MSW solution of the solar neutrino problem}.
\newblock {\em Nucl. Phys. B}, 648:376--396, 2003.

\bibitem{Aboubrahim:2013yfa}
Amin Aboubrahim, Tarek Ibrahim, Ahmad Itani, and Pran Nath.
\newblock {Large Neutrino Magnetic Dipole Moments in MSSM Extensions}.
\newblock {\em Phys. Rev. D}, 89(5):055009, 2014.

\bibitem{Fukuyama:2003uz}
Takeshi Fukuyama, Tatsuru Kikuchi, and Nobuchika Okada.
\newblock {Neutrino magnetic moments and minimal supersymmetric SO(10) model}.
\newblock {\em Int. J. Mod. Phys. A}, 19:4825--4834, 2004.

\bibitem{Healey:2013vka}
Kristopher~J. Healey, Alexey~A. Petrov, and Dmitry Zhuridov.
\newblock {Nonstandard neutrino interactions and transition magnetic moments}.
\newblock {\em Phys. Rev. D}, 87(11):117301, 2013.
\newblock [Erratum: Phys.Rev.D 89, 059904 (2014)].

\bibitem{Papoulias:2015iga}
D.~K. Papoulias and T.~S. Kosmas.
\newblock {Neutrino transition magnetic moments within the non-standard
  neutrino\textendash{}nucleus interactions}.
\newblock {\em Phys. Lett. B}, 747:454--459, 2015.

\bibitem{PhysRevD.103.095004}
Oleg~G. Kharlanov and Pavel~I. Shustov.
\newblock Effects of nonstandard neutrino self-interactions and magnetic moment
  on collective majorana neutrino oscillations.
\newblock {\em Phys. Rev. D}, 103:095004, May 2021.

\bibitem{Capozzi:2020cbu}
Francesco Capozzi and Georg Raffelt.
\newblock {Axion and neutrino bounds improved with new calibrations of the tip
  of the red-giant branch using geometric distance determinations}.
\newblock {\em Phys. Rev. D}, 102(8):083007, 2020.

\bibitem{PhysRevLett.111.231301}
N.~Viaux, M.~Catelan, P.~B. Stetson, G.~G. Raffelt, J.~Redondo, A.~A.~R.
  Valcarce, and A.~Weiss.
\newblock Neutrino and axion bounds from the globular cluster m5 (ngc 5904).
\newblock {\em Phys. Rev. Lett.}, 111:231301, Dec 2013.

\bibitem{2015APh....70....1A}
S.~{Arceo-D{\'\i}az}, K.~P. {Schr{\"o}der}, K.~{Zuber}, and D.~{Jack}.
\newblock {Constraint on the magnetic dipole moment of neutrinos by the tip-RGB
  luminosity in {\ensuremath{\omega}}-Centauri}.
\newblock {\em Astroparticle Physics}, 70:1--11, October 2015.

\bibitem{Okun:1986hi}
L.~B. Okun, M.~B. Voloshin, and M.~I. Vysotsky.
\newblock {Electromagnetic Properties of Neutrino and Possible Semiannual
  Variation Cycle of the Solar Neutrino Flux}.
\newblock {\em Sov. J. Nucl. Phys.}, 44:440, 1986.

\bibitem{Nunokawa:1993dr}
H.~Nunokawa and H.~Minakata.
\newblock {Neutrino spin precession with flavor mixing and the solar neutrino
  problem}.
\newblock In {\em {28th Rencontres de Moriond: Perspectives in Neutrinos,
  Atomic Physics and Gravitation}}, pages 161--164, 1993.

\bibitem{joshi2020neutrino}
Sandeep Joshi and Sudhir~R Jain.
\newblock Neutrino spin-flavor oscillations in solar environment.
\newblock {\em Research in Astronomy and Astrophysics}, 20(8):123, 2020.

\bibitem{PhysRevD.37.1368}
Chong-Sa Lim and William~J. Marciano.
\newblock Resonant spin-flavor precession of solar and supernova neutrinos.
\newblock {\em Phys. Rev. D}, 37:1368--1373, Mar 1988.

\bibitem{Raffelt:1999tx}
Georg~G. Raffelt.
\newblock {Particle physics from stars}.
\newblock {\em Ann. Rev. Nucl. Part. Sci.}, 49:163--216, 1999.

\bibitem{Lychkovskiy:2009pm}
Oleg Lychkovskiy and Sergei Blinnikov.
\newblock {Spin flip of neutrinos with magnetic moment in core-collapse
  supernova}.
\newblock {\em Phys. Atom. Nucl.}, 73:614--624, 2010.

\bibitem{Vassh:2015yza}
N.~Vassh, E.~Grohs, A.~B. Balantekin, and G.~M. Fuller.
\newblock {Majorana Neutrino Magnetic Moment and Neutrino Decoupling in Big
  Bang Nucleosynthesis}.
\newblock {\em Phys. Rev. D}, 92(12):125020, 2015.

\bibitem{popov2019neutrino}
Artem Popov and Alexander Studenikin.
\newblock Neutrino eigenstates and flavour, spin and spin-flavour oscillations
  in a constant magnetic field.
\newblock {\em The European Physical Journal C}, 79(2):1--7, 2019.

\bibitem{sasaki2021neutrino}
Hirokazu Sasaki and Tomoya Takiwaki.
\newblock Neutrino-antineutrino oscillations induced by strong magnetic fields
  in dense matter.
\newblock {\em Physical Review D}, 104(2):023018, 2021.

\bibitem{2020NatAs...4..690P}
Steven~G. {Parsons}, Alexander~J. {Brown}, Stuart~P. {Littlefair}, Vikram~S.
  {Dhillon}, Thomas~R. {Marsh}, J.~J. {Hermes}, Alina~G. {Istrate}, Elm{\'e}
  {Breedt}, Martin~J. {Dyer}, Matthew~J. {Green}, and David~I. {Sahman}.
\newblock {A pulsating white dwarf in an eclipsing binary}.
\newblock {\em Nature Astronomy}, 4:690--696, March 2020.

\bibitem{1983bhwd.book.....S}
Stuart~L. {Shapiro} and Saul~A. {Teukolsky}.
\newblock {\em {Black holes, white dwarfs, and neutron stars : the physics of
  compact objects}}.
\newblock 1983.

\bibitem{1969ApJ...155..221S}
M.~P. {Savedoff}, H.~M. {van Horn}, and S.~C. {Vila}.
\newblock {Late Phases of Stellar Evolution. I. Pure Iron Stars}.
\newblock {\em \apj}, 155:221, January 1969.

\bibitem{1969ApJ...156.1021K}
G.~S. {Kutter} and M.~P. {Savedoff}.
\newblock {Evolution of Initially Pure \^\{12\}C Stars and the Production of
  Planetary Nebulae}.
\newblock {\em \apj}, 156:1021, June 1969.

\bibitem{1966ApJ...146..437V}
Samuel~C. {Vila}.
\newblock {Pre-White-Dwarf Evolution. I}.
\newblock {\em \apj}, 146:437, November 1966.

\bibitem{Bedaque:2012mr}
Paulo~F. Bedaque, Evan Berkowitz, and Aleksey Cherman.
\newblock {Neutrino Emission from Helium White Dwarfs with Condensed Cores}.
\newblock 3 2012.

\bibitem{2016ApJ...824...14C}
Matteo {Cantiello}, Jim {Fuller}, and Lars {Bildsten}.
\newblock {Asteroseismic Signatures of Evolving Internal Stellar Magnetic
  Fields}.
\newblock {\em \apj}, 824(1):14, June 2016.

\bibitem{2015Sci...350..423F}
Jim {Fuller}, Matteo {Cantiello}, Dennis {Stello}, Rafael~A. {Garcia}, and Lars
  {Bildsten}.
\newblock {Asteroseismology can reveal strong internal magnetic fields in red
  giant stars}.
\newblock {\em Science}, 350(6259):423--426, October 2015.

\bibitem{Drewes:2021fjx}
Marco Drewes, Jamie McDonald, Lo\"\i{}c Sablon, and Edoardo Vitagliano.
\newblock {Neutrino Cooling Bounds on the Internal Magnetic Fields of White
  Dwarfs}.
\newblock 9 2021.

\bibitem{Fraija:2019qpp}
Nissim Fraija, Enrique~Moreno M\'endez, Gibr\'an Morales, and Alfredo Saracho.
\newblock {Neutrino Signal from Compact Objects during their Formation, their
  Mergers, or as a Signature of Electric-Charge Phase Transition}.
\newblock 5 2019.

\bibitem{Abe:2011ts}
K.~Abe et~al.
\newblock {Letter of Intent: The Hyper-Kamiokande Experiment --- Detector
  Design and Physics Potential ---}.
\newblock 9 2011.

\bibitem{LENA:2011ytb}
Michael Wurm et~al.
\newblock {The next-generation liquid-scintillator neutrino observatory LENA}.
\newblock {\em Astropart. Phys.}, 35:685--732, 2012.

\bibitem{JUNO:2015zny}
Fengpeng An et~al.
\newblock {Neutrino Physics with JUNO}.
\newblock {\em J. Phys. G}, 43(3):030401, 2016.

\bibitem{Esteban:2018azc}
Ivan Esteban, M.~C. Gonzalez-Garcia, Alvaro Hernandez-Cabezudo, Michele
  Maltoni, and Thomas Schwetz.
\newblock {Global analysis of three-flavour neutrino oscillations: synergies
  and tensions in the determination of $\theta_{23}$, $\delta_{CP}$, and the
  mass ordering}.
\newblock {\em JHEP}, 01:106, 2019.

\end{thebibliography}

\end{document}